%% file: xspec.tex
\documentstyle{mn}
\input{epsf_mn}

\def\figinsert#1#2{\epsfbox{#1} \message{#2} }    
\begin{document}
\title[A deep $\em ROSAT$ survey - XII. The X-ray spectra 
 of faint $\em ROSAT$  sources]
{A deep $\em ROSAT$ survey - XII. 
The X-ray spectra of faint {\it ROSAT} sources}
\author[O. Almaini et al.]
{O.~Almaini$^1$\thanks{Present address: Institute Of Astronomy, Madingley Road, Cambridge CB3 OHA}, 
T.~Shanks$^1$,
B.J.~Boyle$^2$,
R.E.~Griffiths$^3$,  
N. ~Roche$^3$, \cr
G.C.~Stewart$^4$, 
and I.~Georgantopoulos$^4$ \\
$^1$Department of Physics, University of Durham, South Road,
Durham, DH1 3LE, UK.\\
$^2$Royal Greenwich Observatory, Madingley Road, Cambridge, 
CB3 OEZ, UK. \\
$^3$Johns Hopkins University, Homewood Campus, 
Baltimore MA21260, USA. \\
$^4$Department of Physics, University of Leicester, University of 
Leicester, LE1 7RH, UK. }

\date{Accepted May 16th 1996. Submitted November 14th 1995.}

\maketitle

\begin{abstract}

Optical spectroscopy has enabled us to identify the optical
counterparts to over 200 faint X-ray sources to a flux limit of
$S_{(0.5-2keV)}=4\times10^{-15}$erg$\,$s$^{-1}$cm$^{-2}$ on 5 deep
$\em ROSAT$ fields.  Here we present a spectral analysis of all the
X-ray sources to investigate claims that the average source spectra
harden at faint X-ray flux. From a hardness ratio analysis we confirm
that the average spectra from $0.5-2$\,keV harden from an equivalent
photon index of $\Gamma=2.2$ at
$S_{(0.5-2keV)}=1\times10^{-13}$erg$\,$s$^{-1}$cm$^{-2}$ to
$\Gamma\simeq1.7$ below
$1\times10^{-14}$erg$\,$s$^{-1}$cm$^{-2}$. These spectral changes are
due to the emergence of an unidentified source population rather than
the class of X-ray QSOs already identified. The 128 QSOs detected so
far show no evidence for spectral hardening over this energy range and
retain a mean photon index of $\Gamma=2.2$.  Recent work suggests that
many of the remaining unidentified sources are X-ray luminous
galaxies. Taking a subset identified as the most likely galaxy
candidates we find that these show significantly harder spectra than
QSOs.  The emission line galaxies in particular show spectra more
consistent with the residual X-ray background, with $\Gamma=1.51 \pm
0.1$ from $0.1-2\,$keV.  Individually the galaxies appear to be a mixture of
absorbed and unabsorbed X-ray sources. Combined with recent
cross-correlation results and work on the source number count
distribution, these results  suggest that we may be uncovering the
missing hard component of the cosmic X-ray background.

\end{abstract}


\section{Introduction}

We are conducting a survey to understand the nature of the faint X-ray 
sources identified on deep ($21-49\,$ks)  {\it ROSAT} PSPC exposures. 
So far we have identified over 100 QSOs from 5 $\it ROSAT$ fields
and shown that QSOs make up at least $\sim30\%$
of the X-ray background (XRB) at 1keV (Shanks et al 1991).
However, studies of the QSO X-ray luminosity function (Boyle et al 1994) and 
the number count distribution (Georgantopoulos et al 1995) suggest that 
the known QSO population is
unlikely to form more than $50\%$ of the total XRB flux.
QSOs also show  relatively steep X-ray spectra
with  indices of $\Gamma = 2.2\pm 0.1$ while 
the extragalactic XRB from 1-10 keV has a flatter power-law 
index of $\Gamma = 1.4$ (Gendreau et al 1995). 
This suggests that we need a new, faint source population with a flatter 
X-ray spectrum  to account for the remainder of the  background radiation.

From these deep {\it ROSAT} exposures it is also beginning to emerge
that many of the remaining X-ray sources are associated with faint
galaxies. These appear to be a mixture of absorption and emission line
galaxies with optical spectra and redshifts typical of the galaxy
population, but the implied X-ray luminosities are $10-100$ times
higher than those of similar galaxies locally (Roche at al 1995a,
Boyle et al 1995a, Griffiths et al 1995a, 
Carballo et al 1995, McHardy et al 1995).  The
nature of the X-ray emission mechanism in these galaxies is still not
clear, but recent work at brighter flux limits (Boyle et al 1995b)
suggests that some may be Seyfert 2 or starburst galaxies. The
clearest evidence that faint galaxies are significant contributors to
the XRB has come from the spatial cross-correlation of XRB
fluctuations and faint $B<23$ galaxies (Roche et al 1995a). This
statistical method avoids the source confusion problem that prevents
faint galaxies from being associated with X-ray sources.  The
amplitude of the cross-correlation implies that $B<23$ galaxies
directly contribute some 17$\pm$2\% of the 1keV XRB.  Integrating the
implied local X-ray volume emissivity to faint magnitudes and high
redshifts suggests that the remainder of the soft XRB can be explained
by faint galaxies (Roche et al 1995a, Almaini 1996). Their potential
contribution to the hard XRB depends critically on their X-ray spectra.

In this paper we investigate the X-ray spectra of all the sources
identified on 5 deep {\it ROSAT} fields. 
Recent work by Carballo et al (1995) has suggested that X-ray luminous 
galaxies show
harder spectra than QSOs.
Other deeper surveys (Hasinger
et al 1993, Vikhlinin et al 1994) have revealed the possibility that
the mean spectra of the source population may harden as we go to fainter
X-ray fluxes, perhaps indicating that we are beginning to identify the
missing faint sources required to explain the remainder of the X-ray
background. In this paper we attempt to identify the type of source
responsible for this trend. In section 2 we present our data set and
the data reduction techniques and in section 3 we use model
independent hardness ratios to investigate the X-ray spectra of the
QSOs and other source types. In section 4 we repeat this analysis with
full spectral fitting followed by our conclusions and a discussion in
section 5.

\section{Observational data }

\subsection{The sample}

Here we use 5 deep ($21-49\,$ks) pointed observations with the {\it
ROSAT} PSPC with optical identifications from the X-ray source
catalogue of Shanks et al 1996 (in preparation).  These are well
studied optical fields selected from the ultra-violet excess (UVX)
survey of Boyle et al (1990). Our analysis is restricted to the
central 18 arcminute radius of the {\it ROSAT} pointings to maximise
the sensitivity of our observations since the point spread function of
the PSPC rapidly increases beyond the central 20 arcminute radius. Due
to the considerable contamination from both the galactic background
and solar scattered X-rays below 0.5\,keV (Snowden \& Freyberg 1993)
we optimize the sensitivity of source detection by concentrating on
the 0.5-2.0\,keV data.

Full details of the X-ray source detections and optical spectroscopic
identifications will be given elsewhere (Shanks et al, in preparation)
and so only brief details will be given below.  Sources were
identified using the standard PSS algorithm within the $\it ASTERIX$
data processing package, which detects peaks above a certain threshold
and matches the expected PSF to the background fluctuations to
determine whether the source is real. In this way, 356 X-ray sources
were detected above a 4$\sigma$ significance and 197 sources were
detected above 5$\sigma$ in the 0.5-2.0\,keV band over 5 {\it ROSAT}
fields.  Optical counterparts to these X-ray sources were identified
from COSMOS and APM measurements of J and U band UK Schmidt
plates. Astrometric transforms between {\it ROSAT}\, X-ray and
COSMOS/APM co-ordinates were set up using the Durham/AAT UVX QSOs
detected by {\it ROSAT} on each field.  Low resolution (12\AA) optical
spectra were then obtained for the nearest optical counterpart to each
X-ray source using the AUTOFIB multi-object system at the
Anglo-Australian Telescope.  A summary of the optical identifications
of the 4$\sigma$ sources is given in Table 1. Note that this is
considerably less complete than the identifications of the smaller
list of 5$\sigma$ sources listed in Georgantopoulos et al (1996) since
in this work we are attempting to probe fainter flux limits.  Of the
257 sources for which optical identifications were attempted, 128 were
identified as QSOs and Seyfert 1 galaxies which directly account for
$\sim$30$\%$ of the total XRB at 1keV (Shanks et al 1991).  Less than
10$\%$ of the sources were found to be galactic late type stars. Of
the remaining positive identifications, 10 continuum objects and the
emission from a galaxy cluster were also detected (Roche et al 1995b).
However, as can be seen from Table 1, a large fraction of the sources
remain unidentified or unobserved (both hereafter referred to as the
``unidentified'' sources).  In many cases, observing limitations
prevented the object from being observed or the S/N in the optical
spectra was too poor to allow a reliable identification.
Interestingly however, $\sim$100 of these unidentified X-ray sources
appeared to be associated with faint, ``normal'' galaxies on
photographic plates and for 38 of these sources the optical
counterpart was firmly identified as a galaxy by spectroscopy.
However, due to the high sky density of galaxies at faint magnitudes
($\sim$10000 deg$^{-2}$ at B$<$23, Metcalfe et al 1991) and the
$\sim$25$''$ FWHM X-ray error circle many of these will be chance
coincidences. A reliable estimate of the contribution of faint
galaxies to the XRB can only be determined statistically (see Roche et
al 1995a). For the galaxies at brighter limiting magnitudes the
confusion problem becomes less pronounced. We therefore identify a
sample of ``probable'' galaxy candidates with B$<$21.5 for which the
optical counterpart lies within 20$''$ of the X-ray source.  This
optical magnitude represents the limit of reliable galaxy
identification on a UK Schmidt plate. Cross-correlating COSMOS and APM
galaxy catalogues to the same magnitude limit with the unidentified
X-ray sources we estimate that $\sim$6 of this restricted sample will
be spurious identifications.  15 of these galaxies were identified
with narrow emission line features and 8 were identified as absorption
line galaxies.  The sample used by Griffiths et al 1995b is taken from
the 5$\sigma$ subset of these galaxies.  Further details of the
properties of these sources are given in Table 5 and in Shanks et al
1996.

\begin{figure}
\centering \centerline{\epsfxsize=8truecm
\figinsert{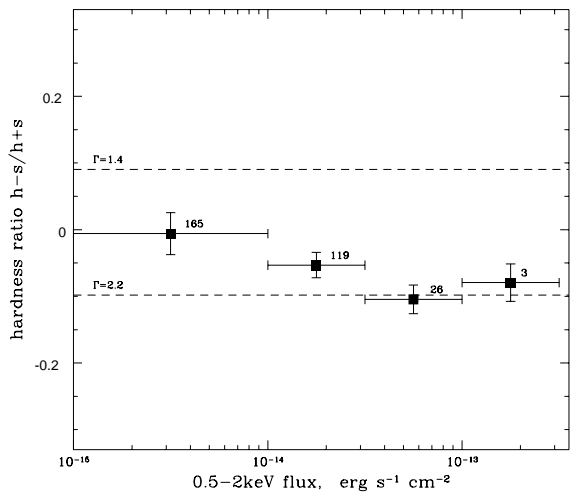}{0.0pt}}
\caption{$0.5-2$\,keV hardness ratios 
for the stacked spectra of 313 X-ray sources from 5 deep  $\em ROSAT$ fields
binned according to flux.  
The number of X-ray sources within each bin is indicated. For comparison, the
hardness ratios for 2 power law models are also shown.}
\end{figure}

\begin {table}
\begin {center}
\caption {Summary of optical identifications to 4$\sigma$ X-ray sources from
5 deep $\em ROSAT$ fields.}
\begin {tabular}{||c|c||}
AGN  & 128 \\
Stars  & 27 \\
Continuum & 10 \\
Clusters & 1 \\
Unobserved & 96  \\ 
Unidentified & 89 \\
(Probable galaxies) & (23) \\
\hline
Total & 356 \\
\hline
\end{tabular}
\end{center}
\end{table}

\subsection{Obtaining X-ray spectra}

For each source the X-ray counts used to determine fluxes, hardness
ratios and spectra were obtained using a circle that encloses 90$\%$
of the source photons. The radius of this circle varies with energy
and the off axis angle (Hasinger et al 1992).  Data from periods of
high particle background were excluded from the analysis, excluding
approximately 10$\%$ of the data when the Master Veto Rate was above
170 counts s$^{-1}$ (Plucinsky et al 1993).  Due to the faint nature
of many of these sources, considerable care was taken in choosing an
area for background subtraction. Possible problems include
irregularities in the galactic background or contamination from solar
scattered X-rays.  However, after the subtraction of sources the
residual background levels were found to remain constant over the 18
arcminute central region and no significant gradient was apparent on
any field. Circular areas of 4 to 6 arcminute radius were then chosen
from source free regions to perform the background subtraction,
correcting for the vignetting between source and background boxes.

\section{Hardness ratios}

\subsection{QSOs and unidentified sources}

Since the majority of our sources have fewer than 100 total counts in
the $\em ROSAT$ band, detailed spectral fitting is not possible. We
therefore derive model independent hardness ratios to compare the
spectral properties of these sources. By forming a ``soft'' energy
band ($S$) from the $0.5-1$\,keV flux and a ``hard'' band ($H$) from
$1-2$\,keV we define the hardness ratio as;

\begin{equation}
HR=\frac{H-S}{H+S}
\end{equation}
 
As explained above, 
our sample was initially selected by excluding the data below 0.5 keV
 to allow a higher efficiency in source detection. 
We therefore ignore the very soft flux in the first instance
 and  define our hardness ratios from $0.5-2$\,keV in order to 
characterize the source population fairly without a preferential 
selection of hard sources.

To test for possible systematic biases that might arise due to the
combined energy and radial dependence of the PSF the entire sample was
split into sources lying within a 10 arminute radius from the centre
of the PSPC and those lying beyond.  No trend in hardness ratios with
off-axis angle was apparent at any flux, as verified by a Kolmogorov
Smirnov test on the data. Another potential problem in analysing $\em
mean$ hardness ratios would be an artificial skewness in the
distribution at faint X-ray flux. If the instrument is more sensitive
in either the $H$ or $S$ bands, individual hardness ratios may be
skewed towards $+1$ or $-1$ as the flux tends to zero and becomes
dominated by noise. To overcome this, we will plot hardness ratios of
the $\em stacked$ spectra in each flux bin. However, the similarity of these
distributions to those obtained with mean hardness ratios
suggest that this problem does not significantly affect our data.

Removing the known galactic stars, BL Lac candidates and the cluster
emission\footnote{Note that replacing these sources has a negligible
affect on any of the results presented here.}, we plot the stacked
hardness ratios for the other 313 X-ray sources on Figure 1, binned
as a function of flux.  These results show a hardening of
the mean source spectra with decreasing flux, as previously suggested
by Hasinger et al (1993) and Vikhlinin (1994).

\begin{figure}
\centering
\centerline{\epsfxsize=8truecm 
\figinsert{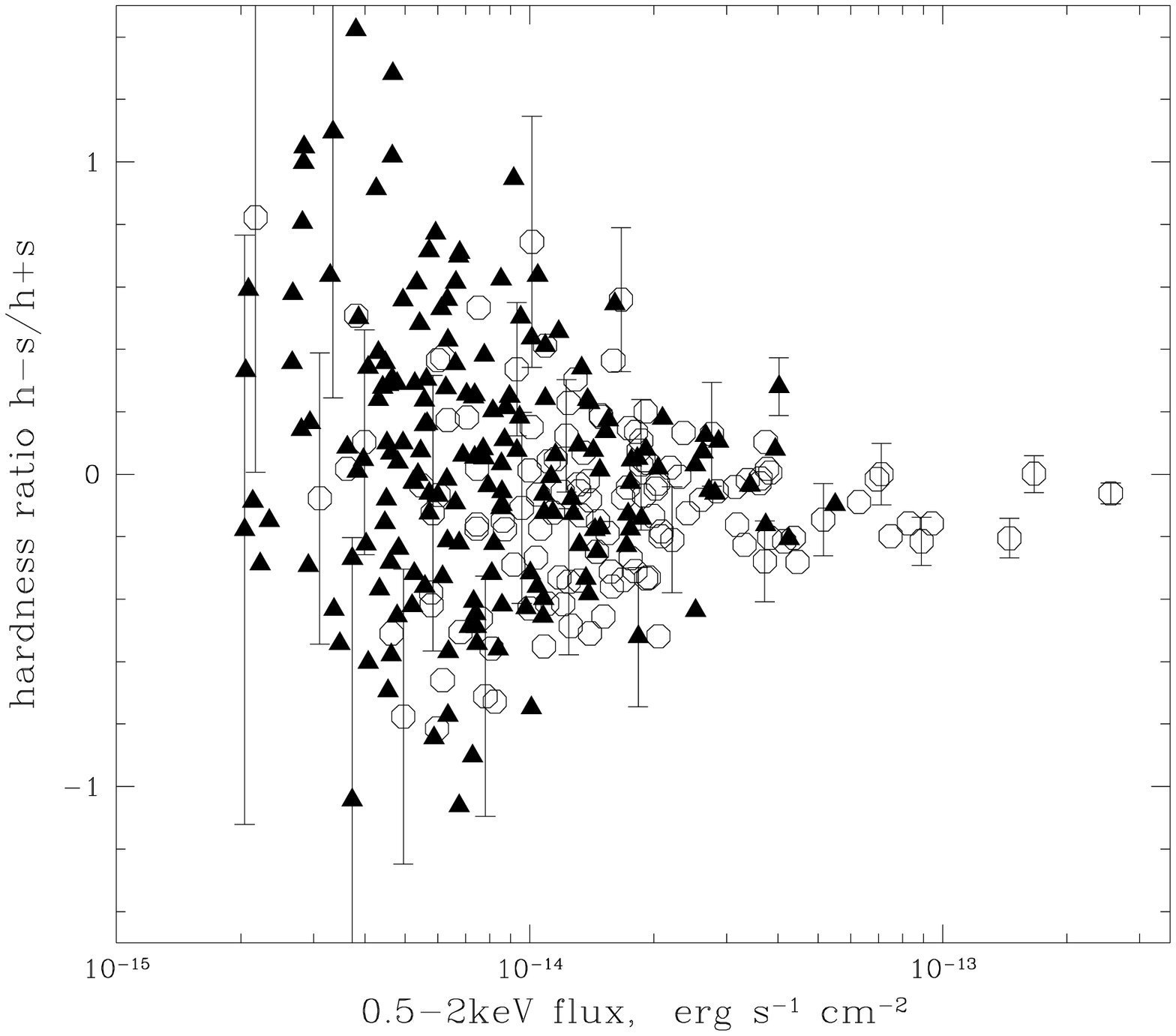}{0.0pt}}
\centerline{\epsfxsize=8truecm 
\figinsert{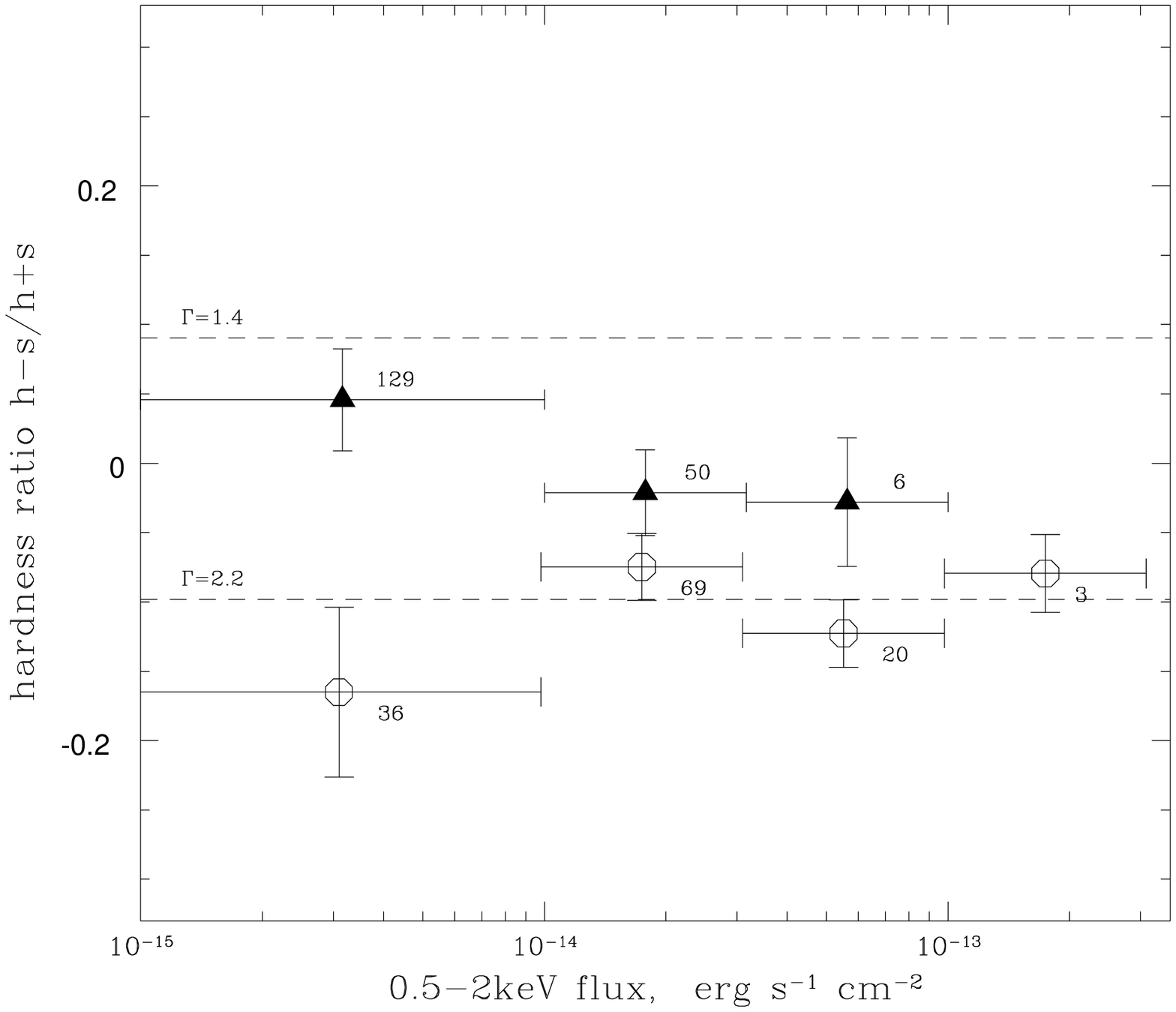}{0.0pt}}
\caption{(a) Individual $0.5-2$\,keV hardness ratios as a function of
flux for QSOs (unfilled circles) and unidentified X-ray sources and
galaxies (filled triangles).  For clarity, the appropriate 1$\sigma$
errors are only displayed for a representative selection of sources.
On (b) we plot hardness ratios for stacked spectra binned according to
flux, as on Figure 1 , but separating the QSOs from the unidentified
sources and galaxies.  Note the change in scale compared to (a).}
\end{figure}

\begin{figure}
\centering
\centerline{\epsfxsize=8truecm 
\figinsert{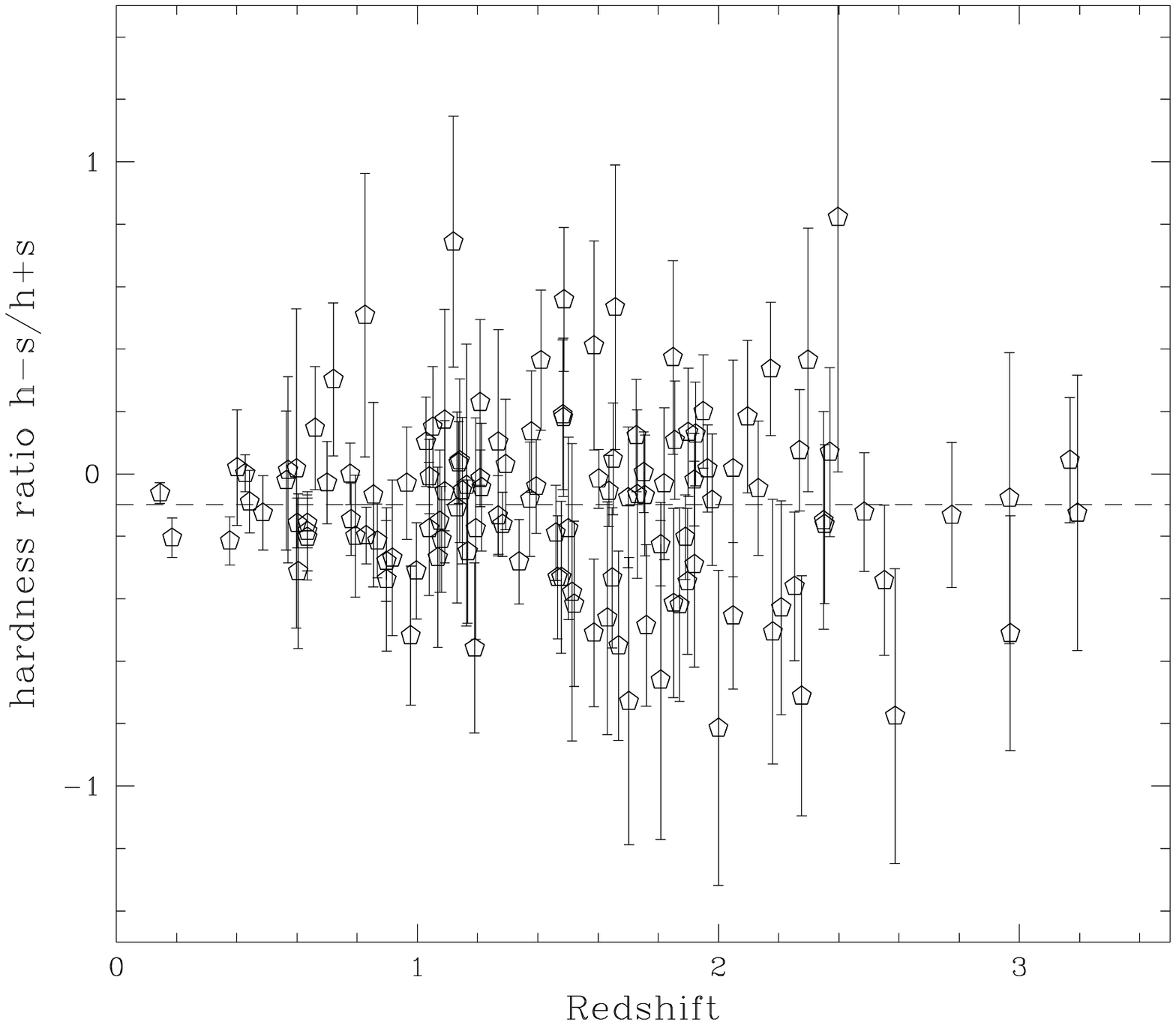}{0.0pt}}
\centering
\centerline{\epsfxsize=8truecm 
\figinsert{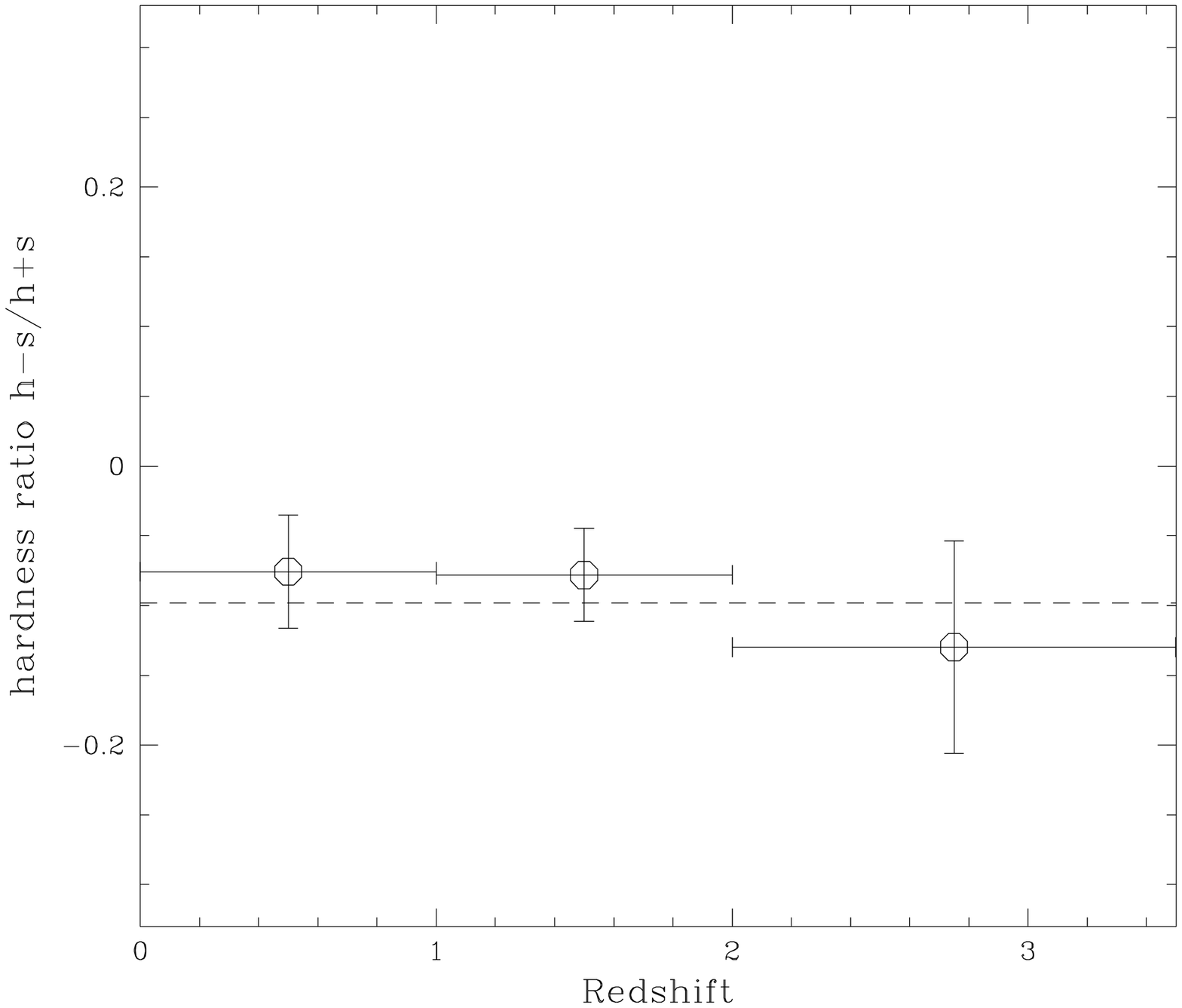}{0.0pt}}
\caption{(a) Showing hardness ratios as a function of redshift for the 
128 QSOs detected on these fields, while  (b) shows the stacked
hardness ratios when binned according to redshift with 1$\sigma$
errors representing the rms error on the mean. The dotted line
shows the mean QSO hardness ratio.}
\end{figure}

Since AGN are the main contributors to the total source flux at
brighter energies (eg. Shanks et al 1991), there have been suggestions
that an evolution in AGN X-ray spectra may be responsible for the
trend in hardness ratios.
A hardening of QSO spectra towards higher redshift has been
postulated, due to either a change in the actual intrinsic spectrum
(Morisawa et al  1995) or the effect of intervening absorption
from damped Ly$\alpha$ systems (Vikhlinin et al 1995).  On Figure 2(a)
we plot the individual hardness ratios, separating QSOs from the
other, mostly unidentified sources.  The dominant feature on this
diagram is the large spread in hardness ratios towards fainter fluxes
due to counting statistics.  We therefore bin these hardness ratios
according to flux, and Figure 2(b) displays the hardness ratios for
the stacked spectra in 4 flux bins.

Several features are immediately apparent from these
distributions. They show that the unidentified sources have harder
mean X-ray spectra than QSOs, regardless of source intensity.  A
Kolmogorov-Smirnov test yields a $>99.9 \%$ probability that the two
distributions shown on Figure 2(a) do not arise from the same parent
population. This is consistent with the errors on the stacked spectra
shown on Figure 2(b).  Secondly, the {\em QSOs show no evidence for
spectral hardening with decreasing flux}, indicating that the change
in mean source spectra is due to the emergence of another population
from within the unidentified sources with a harder spectrum than QSOs.
Given the incomplete spectroscopic identification in our survey, the
unidentified population almost certainly contains some contribution
from steep spectrum QSOs. The mean spectrum of the remaining
population may therefore be even harder than indicated on Figure 2. It
is also worth noting that the total spectrum of the unidentified
sources harden below $1\times10^{-14}$erg$\,$s$^{-1}$cm$^{-2}$.  This
may in part be due to a decreasing fractional contamination by
unidentified QSOs, but it is interesting to note that this behaviour
is predicted by models which explain the residual XRB using a
population of sources with curved X-ray spectra (Boyle 1996). The
increased contribution from high redshift objects would produce a
steep log $N$-log $S$ relationship for the unidentified population and
harden the mean spectra at faint fluxes.

\begin{figure}
\centering
\centerline{\epsfxsize=8truecm 
\figinsert{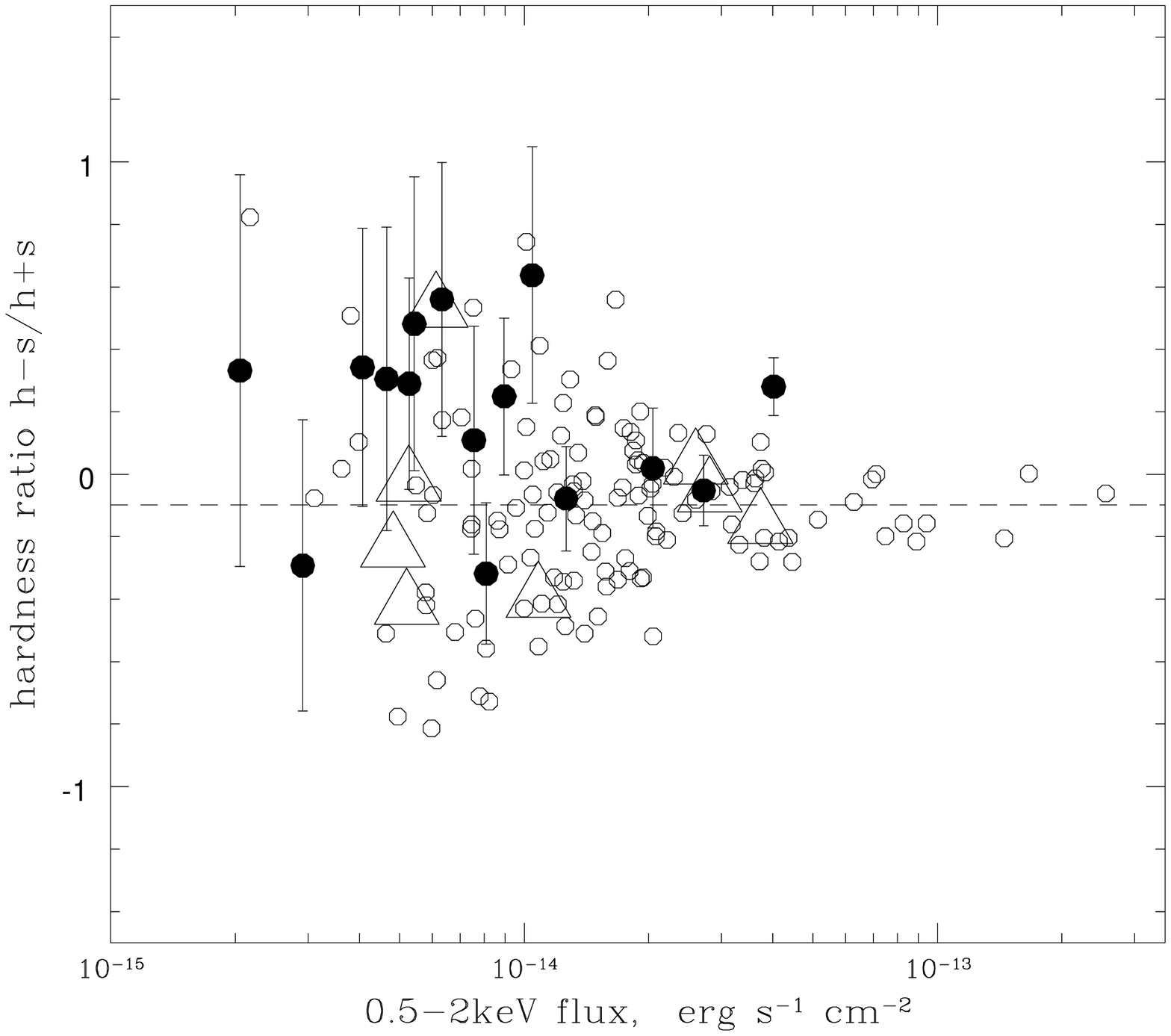}{0.0pt}}
\centering
\centerline{\epsfxsize=8truecm 
\figinsert{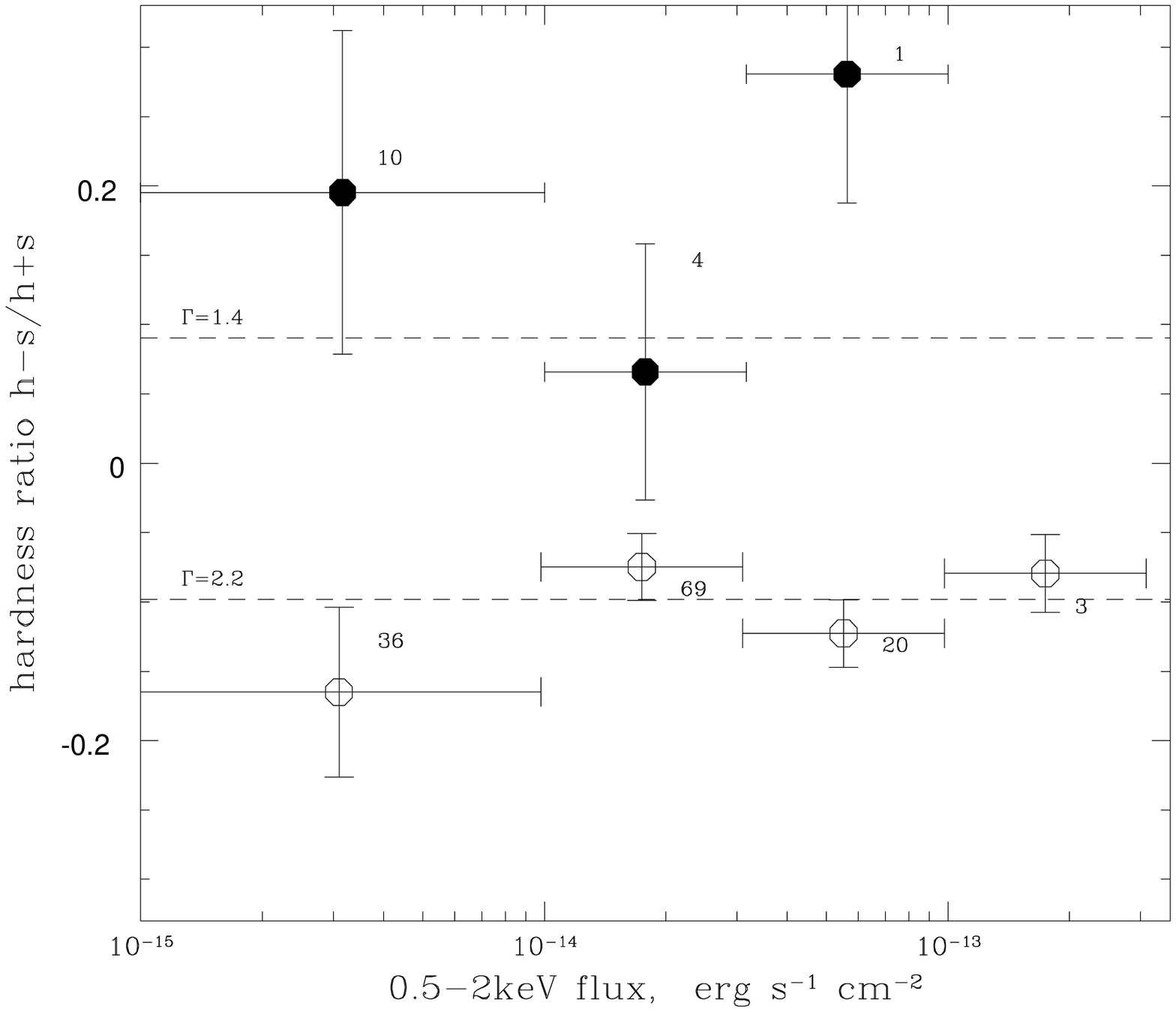}{0.0pt}}
\caption{(a) Hardness ratios as a function of flux for the 
23 most likely  X-ray emitting galaxies with the sample of 128 QSOs 
(unfilled circles) for comparison. Galaxies are separate into 8 absorption
line objects (unfilled triangles) and 15 narrow 
emission line galaxies (filled circles).
For clarity the 
1$\sigma$ error bars are only displayed for the emission-line galaxies.
On (b) we show the hardness ratios for the stacked spectra of the 
emission line galaxies and QSOs as a function of flux.}
\end{figure}

\begin{figure}
\centering
\centerline{\epsfxsize=8truecm 
\figinsert{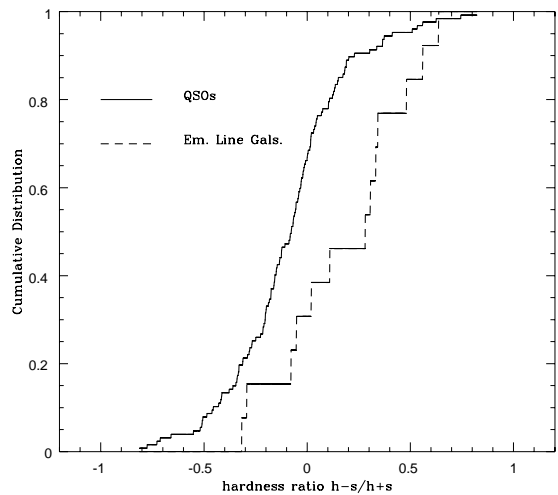}{0.0pt}}
\caption{ Cumulative probability distributions for the 15 emission line 
galaxies and 128 QSOs  shown on Figure 4(a) (Kolmogorov-Smirnov test).}
\end{figure}

On Figures 3(a) and 3(b) we also plot QSO hardness ratios as function
of redshift.  This also illustrates the lack of spectral evolution in
our QSO sample from 0.5-2keV, suggesting that broad-line AGN are
unlikely to account for the missing hard component of the cosmic
XRB. Interestingly however, in the softer band below 0.5keV (where the
cosmic X-ray background is dominated by galactic emission) there {\em
does} appear to be evidence for a change in QSO spectra with redshift.
In this band it is now widely accepted (see Mushotzky et al 1993) that
the spectra of QSOs have a significant contribution from a soft excess
component, generally believed to be thermal emission from an accretion
disk.  Using the same sample of QSOs in an independent analysis,
Stewart et al (1994) find evidence for a hardening in the spectra of
QSOs with redshift in this softer band which has been attributed to
changes in the thermal black-body component (see also Mushotzky et al
1993).  This evolution is due in part to a redshifting of the soft
excess component out of the $\em ROSAT$ passband for higher redshift
QSOs, but there also appears to be evidence for a change in the
temperature and normalisation of this component. However, we are
concerned here with the extragalactic X-ray background above 0.5keV
and in this band there is no evidence for any QSO spectral evolution.

\subsection{X-ray luminous galaxies}

In Section 2 we noted that $\sim$100 of the 185 unidentified sources
appear to be associated with faint optical galaxies.  The
cross-correlation results of Roche et al 1995a suggest that many of
these are likely to be genuine X-ray sources but due to the high sky
density of ``normal'' field galaxies at faint magnitudes there will
also be a significant number of chance associations.  We therefore
selected a restricted sample of the most likely galaxy candidates with
brighter optical magnitudes (B$<$21.5) and lying within 20$''$ of the
X-ray source.  In total, 23 galaxies meet this criteria from which we
expect only $\sim$6 to be spurious identifications (see Section 2.1).
The hardness ratios for these X-ray sources and the 128 QSOs are
displayed on Figure 4(a), separating the 15 narrow emission line
galaxies from the 8 absorption line galaxies.  Despite the limited
sample and the large errors on individual faint sources, there is
clearly evidence that the emission line galaxies in particular come
from a harder population than the QSOs. While the 8 absorption line
galaxies are evenly distributed about the mean hardness ratio for
QSOs, 13 of the 15 emission line galaxies lie formally above this mean
value. On Figure 4 (b) we display the hardness ratios for QSOs and
emission line galaxies binned according to flux.  A Kolmogorov-Smirnov
test yields a 98.6$\%$ probability that the hardness ratios associated
with the emission line galaxies and QSOs do not arise from the same
parent population.  The cumulative distributions are shown on Figure
5.

\newpage 

\section{Spectral fitting}

\subsection{Stacked spectra}

\begin {table}
\begin {center}
\caption {Summary of $\em ROSAT$ fields, with coordinates in J2000 and
galactic column density in $10^{20}\,$atom$\,$cm$^{-2}$.}
\begin {tabular}{||c|c|c|c|c||}
Field & RA & DEC & $N_{H}$ & Exposure time \\
\hline
GSGP4 & 00 57 28.7 & -27 38 24 & 1.8 & 48955 \\
SGP2  & 00 52 04.8 & -29 05 24 & 1.8 & 24494 \\
SGP3  & 00 55 00.0 & -28 19 48 & 1.8 & 21062 \\
QSF1  & 03 42 09.6 & -44 54 36 & 1.7 & 26144 \\
QSF3  & 03 42 14.3 & -44 07 48 & 1.7 & 27358 \\
\hline
\end{tabular}
\end{center}
\end{table}

In Section 3 we used hardness ratios to analyse the X-ray spectra of individual
faint sources. In this section we analyse the X-ray spectra in more detail 
using the full resolution of the PSPC detector by stacking together the 
spectra of different source types. 

Details of the five $\em ROSAT$ fields used in this analysis are
summarised in Table 2.  The column densities of galactic hydrogen are
very similar on each field, but a mean value weighted by exposure
times was used if any stacked spectra were obtained from different
fields.  For the spectral fitting, the response matrix DRM\_06 was
used for observations made before October 1991 (QSF1 and QSF3) while
the matrix DRM\_036 was used for observations made after that date
(SGP2, SGP3 and GSGP4). Although problems in the calibration of the
PSPC (see Turner et al 1995) can lead to some uncertainties in the
spectral fits, we are primarily concerned with broad $\em differences$
between the spectra of different sources which should be unaffected by
these problems.

\begin {table*}
\begin {center}
\caption {Results of power-law fits to the stacked  spectra from all fields, 
separated according to QSOs, unidentified sources and the subset of probable
galaxies. Fits are performed with the full $0.1-2.0$\,keV $\em ROSAT$ 
band and the
restricted $0.5-2.0$\,keV for comparison with hardness ratios. Values of 
photoelectric absorption are fixed at the mean galactic value.}

\begin {tabular}{||c|c|c|c|c|c||}
\hline
  Energy      &  Source Type & No. & $\Gamma$      & $\chi^{2}_{\rm red}$  \\
\hline
 $0.5-2.0$\,keV &  QSOs & 128 &2.23$\pm$0.04 & 0.51 \\
              &  Unidentified  & 185 &1.81$\pm$0.06  & 0.95  \\
	      &  Probable galaxies & 23 &1.64$\pm$0.14  & 1.18  \\
	      &   (Em. line gal.) & 15 &1.36$\pm$0.18  & 1.91  \\
	      &  (Abs. line gal.)& 8 &2.21$\pm$0.23  & 1.25 \\
\hline
 $0.1-2.0$\,keV &  QSOs & 128 & 2.30$\pm$0.01 & 8.70 \\
              &  Unidentified  & 185 &1.74$\pm$0.03  & 3.17  \\
	      &  Probable galaxies & 23 &1.69$\pm$0.06  & 1.45  \\
	      &   (Em. line gal.) & 15 &1.51$\pm$0.09  & 1.79  \\
	      &  (Abs. line gal.) & 8 &1.94$\pm$0.08  & 1.02  \\
\hline
\end{tabular}
\end{center}
\end{table*}

Using the XSPEC spectral analysis  package, we attempt fitting  power-law 
models (modified only by galactic absorption) to the stacked spectra of 
QSOs, unidentified X-ray sources and the subset of probable galaxies.
We emphasize that no particular
physical significance should be attributed to these models and we are merely
attempting to parameterise the  spectral differences between the source types.
For comparison with the hardness
ratio analysis in section 3 we also perform  the fits using only the 
0.5-2keV data. The results  (see Table 3)  confirm
that the unidentified sources, on average, have a harder spectrum
than QSOs. In agreement with the hardness ratio analysis, the subset of 
probable X-ray emitting galaxies (emission line galaxies in particular) 
appear to have a significantly harder spectrum than QSOs. The raw channel 
spectra for the QSOs, 
unidentified sources and the subset of narrow-line X-ray  
 galaxies are displayed on Figure 6 with the best fitting power-law models.

\begin{figure}
\centering
\centerline{\epsfxsize=6truecm 
\figinsert{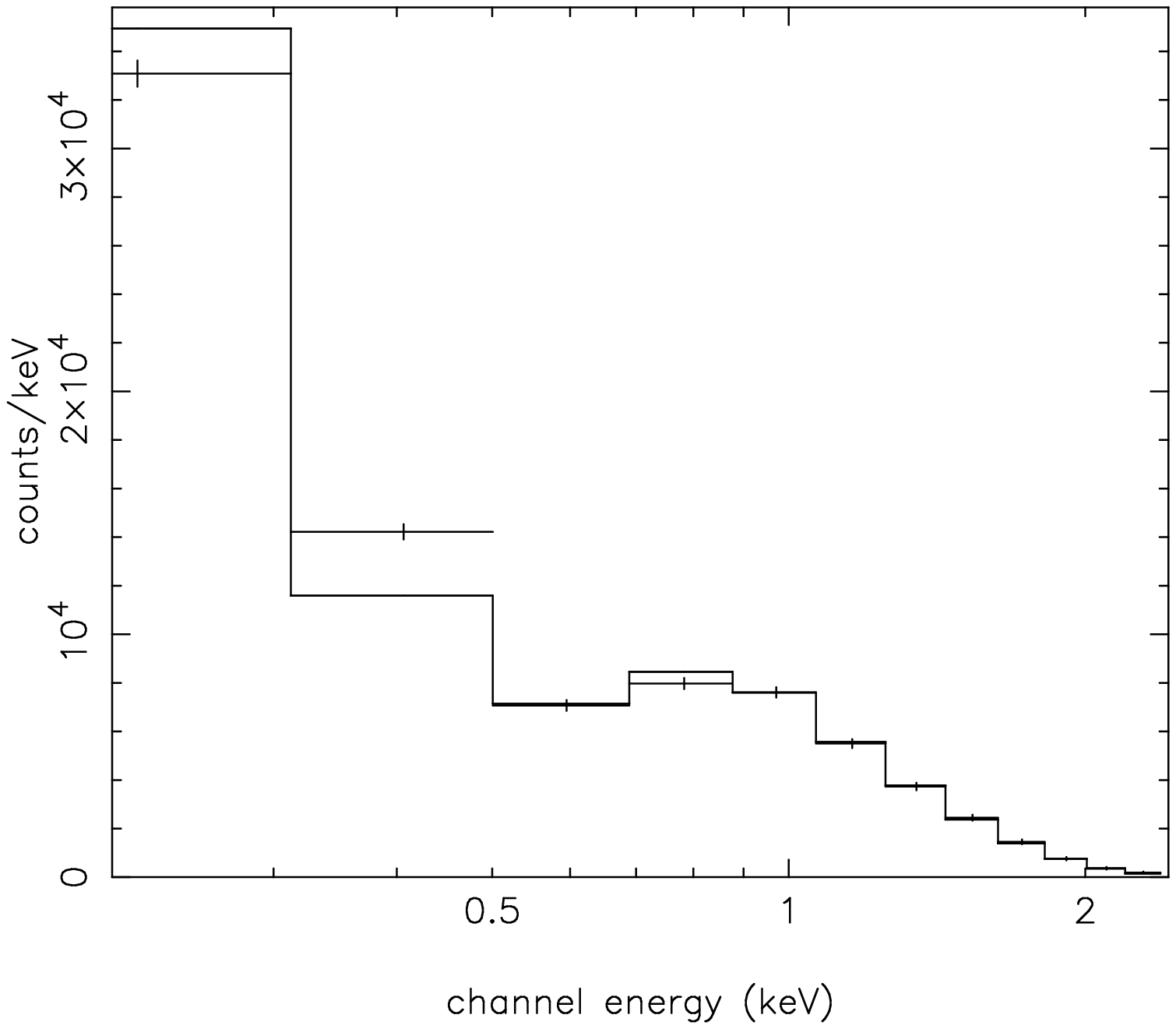}{0.0pt}}
\centering
\centerline{\epsfxsize=6truecm 
\figinsert{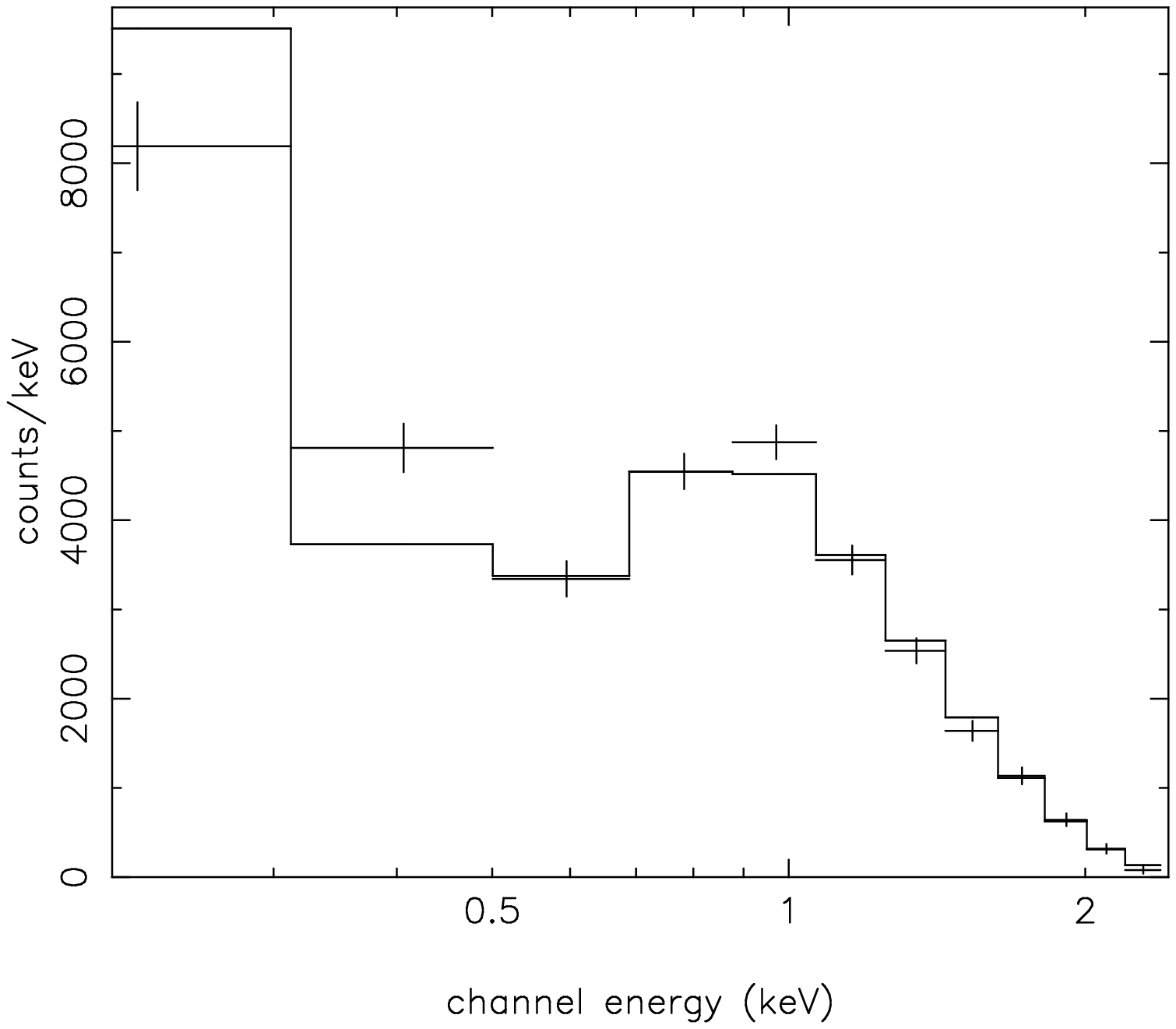}{0.0pt}}
\centering
\centerline{\epsfxsize=6truecm 
\figinsert{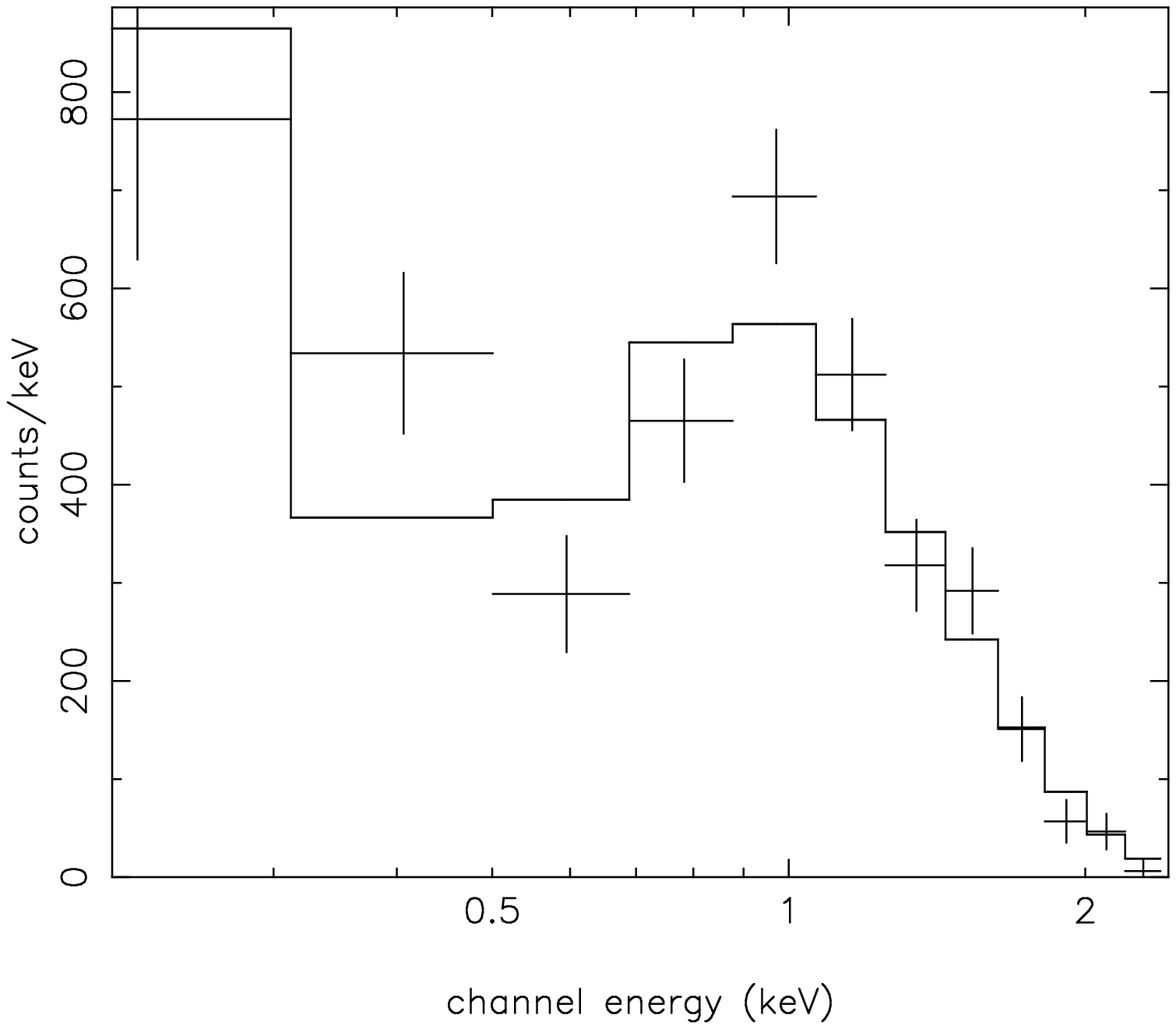}{0.0pt}}
\caption{
The  stacked $0.1-2.0\,$keV  X-ray spectra with  the best fitting power-law 
models for (a) the 128 QSOs  $(\Gamma=2.30\pm0.01)$ , (b) the 
185 unidentified X-ray sources $(\Gamma=1.74\pm0.03)$ and (c) the 15 
probable narrow emission-line galaxies identified as a subset of (b) 
$(\Gamma=1.51\pm0.09)$. In each case  the photoelectric absorption
was fixed at the mean galactic value.}
\end{figure}

Treating each of the 5 $\em ROSAT$ fields separately, Table 4 displays the 
results of power-law fits to the stacked spectra of 
QSOs and unidentified sources. On 4 of 
the 5 fields there are significant spectral  differences between 
the spectra. Note that the largest difference comes from the deepest
(49ks) exposure on the GSGP4 field while on the shortest (21ks) exposure on 
SGP3 the difference in spectra is negligible, consistent with the picture
that a harder population is emerging at fainter fluxes.

As Tables 3 and 4 
show, a simple power-law is a reasonable fit to all the stacked 
data above 0.5keV for all source types. 
For the full band $0.1-2$\,keV 
fits however, the values of $\chi^{2}_{\rm red}$ 
suggest that more detailed  models are required to 
fit the data. For the QSOs in particular, a soft excess component is required
below 0.5keV, as described by Stewart et al (1994),
 but since we 
are interested only in their contribution to the extragalactic XRB we make no 
attempt to model this here. 
Since the  $0.5-2.0$\,keV hardness ratios for QSOs remain constant with 
redshift this would indicate that we are dealing with a power-law
spectrum above $0.5$\,keV without a significant contribution from a soft
excess component.

\begin {table*}
\begin {center}
\caption {Results of power-law fits to the stacked  spectra from each field, 
fixing the photoelectric absorption to the galactic values shown in Table 2.
The fits are also performed using only the $0.5-2.0\,$keV data.}
\begin {tabular}{||c|c|c|c|c|c||}
\hline

              &        &   \multicolumn{2}{c}{QSOs}              & \multicolumn{2}{c}{Unidentified}      \\
       Energy  &  Field &  $\Gamma$      & $\chi^{2}_{\rm red}$   &  $\Gamma$      & $\chi^{2}_{\rm red}$ \\
\hline
 $0.5-2.0$\,keV &  GSGP4 & 2.13$\pm$0.08  & 1.52 & 1.65$\pm$0.11   & 1.08 \\
              &  SGP2  & 2.16$\pm$0.11  & 1.56 & 1.74$\pm$0.15   & 1.01 \\ 
	      &  SGP3  & 2.38$\pm$0.10  & 1.43 & 2.30$\pm$0.13   & 0.89 \\
	      &  QSF1  & 2.19$\pm$0.12  & 1.44 & 1.84$\pm$0.19   & 0.25 \\
	      &  QSF3  & 2.44$\pm$0.11  & 1.98 & 1.71$\pm$0.17   & 1.69 \\
\hline
 $0.1-2.0$\,keV &  GSGP4 & 2.08$\pm$0.03  & 5.63 & 1.36$\pm$0.07  & 3.83 \\
              &  SGP2  & 2.28$\pm$0.03  & 4.65 & 1.63$\pm$0.07  & 2.37 \\
	      &  SGP3  & 2.27$\pm$0.03  & 3.84 & 2.23$\pm$0.04  & 0.96 \\
	      &  QSF1  & 2.41$\pm$0.03  & 2.11 & 1.85$\pm$0.09  & 0.30 \\
	      &  QSF3  & 2.64$\pm$0.03  & 1.98 & 1.95$\pm$0.07  & 1.68 \\
\hline
\end{tabular}
\end{center}
\end{table*}

\newpage

\subsection{Individual galaxy spectra}

The stacked spectra for the X-ray luminous galaxies appear to be
significantly flatter than the combined spectra for the QSOs in our
survey. While most of the individual sources yield a total of fewer
than 40 X-ray photons in the $0.1-2.0$\,keV band, it is important to
establish whether the combined spectrum is due to individual spectra
that are intrinsically flat (Di Matteo \& Fabian 1996) or
alternatively a superposition of absorbed X-ray spectra with
correspondingly distinct low energy cutoffs (Comastri et al 1995, see
also Almaini et al 1995).

In Table 5 we show the results of individual power-law fits (with
galactic absorption) to the X-ray spectra of the 9 brightest 
X-ray emitting galaxies with a $0.5-2.0\,$keV flux 
$>1\times10^{-14}$erg$\,$s$^{-1}$cm$^{-2}$. Spectral
fitting becomes increasingly meangingless for the fainter objects which
were therefore stacked before spectra fitting. Alternatively, equivalent 
photon indices from the hardness ratios on Figure 4 are given
for these fainter galaxies.

A power law model gives a reasonable fit to 7 of the 9 brightest
galaxies and to the stacked spectra of the fainter galaxies. For 2
galaxies however (GSGP4X:091 and GSGP4X:069), simple power-law models
do not give an acceptable fit to the data.  Both of these are emission
line galaxies and they both show very hard X-ray spectra with $\Gamma
< 1$.

\begin{itemize}

\item{{\em\bf GSGP4X:091} For this object, the brightest of the galaxy
candidates, a power-law plus galactic absorption model gives a very
flat $\Gamma$=0.145 but is not a good fit to the data ($\chi^{2}_{\rm
red}$=2.73). A thermal Raymond-Smith model also gives a very poor fit
to the data. The lack of photons at soft energies seems to indicate
photoelectric absorption.  We therefore try adding an absorbing column
at the redshift of the galaxy ($z=0.416$) and assume an intrinsic
power-law of $\Gamma$=2.2 (the mean value for QSOs). This gave a much
improved fit with an intrinsic column density of
$N_H=7.5\pm1.8\times10^{21}\,$atom$\,$cm$^{-2}$ ($\chi^{2}_{\rm
red}$=0.97). The channel spectrum and best fitting model are shown on
Figure 7(a).\footnote{Note that an equally good fit can be obtained
with a similar degree of obscuration and a thermal Raymond-Smith model
with a temperature of $\sim$2keV.}  The X-ray spectrum therefore
presents strong evidence that this is a highly luminous , obscured
X-ray source with an unobscured $0.5-2$\,keV rest-frame luminosity of
$\sim$ $1.3\times 10^{44}\,$erg$\,$s$^{-1}$
($H_0=50\,$km$\,$s$^{-1}$Mpc$^{-1}$, $q_0=0.5$).}

\item{{\em\bf GSGP4X:069} A power-law with only galactic absorption
gives a very flat $\Gamma$=0.79, but this is not a good fit to the
data ($\chi^{2}_{\rm red}$=3.62).  This faint source has only 46
photons from $0.5-2.0$\,keV but nevertheless it also shows evidence
for photoelectric absorption at low energies. Repeating the background
subtraction with various source free regions near the source confirms
that this is not a systematic effect.  Adding an absorbing column at
the redshift of the galaxy ($z=0.213$) and fixing the intrinsic power-law
component to $\Gamma=2.2$ gave a much improved fit to the data
($\chi^{2}_{\rm red}$=0.53) with a restframe absorbing column of
$N_H=2.7\pm1.9\times10^{21}\,$atom$\,$cm$^{-2}$.  The channel spectrum
and best fitting absorbed model are shown on Figure 7(b).}
 
\end{itemize}

While only these 2 galaxies {\em require} additional absorption above
the galactic value, many of the remaining galaxies fit power-law X-ray
spectra significantly flatter than QSOs.  It would be interesting to
determine if these galaxies are intrinsically flat or whether this is
also due to obscuration. Unfortunately there are insufficient photons
to provide useful constraints on both the power-law index and the
rest-frame absorption.  Co-adding the fainter spectra does not resolve
this ambiguity. Stacking the 14 faintest galaxies with flux
$S_{(0.5-2keV)}<1\times10^{-14}$erg$\,$s$^{-1}$cm$^{-2}$, we obtain a
good fit with an unabsorbed power law of index $\Gamma=1.68 \pm 0.11
\, (\chi^{2}_{\rm red}$=0.83).  An equally good fit is obtained with
an intrinsic QSO-like power law ($\Gamma=2.2$) and a low level of
additional photoelectric absorption
($N_H\sim1\times10^{21}\,$atom$\,$cm$^{-2}$). We conclude that further
data are required to differentiate between these possibilities.

\begin{figure}
\centering
\centerline{\epsfxsize=5truecm 
\figinsert{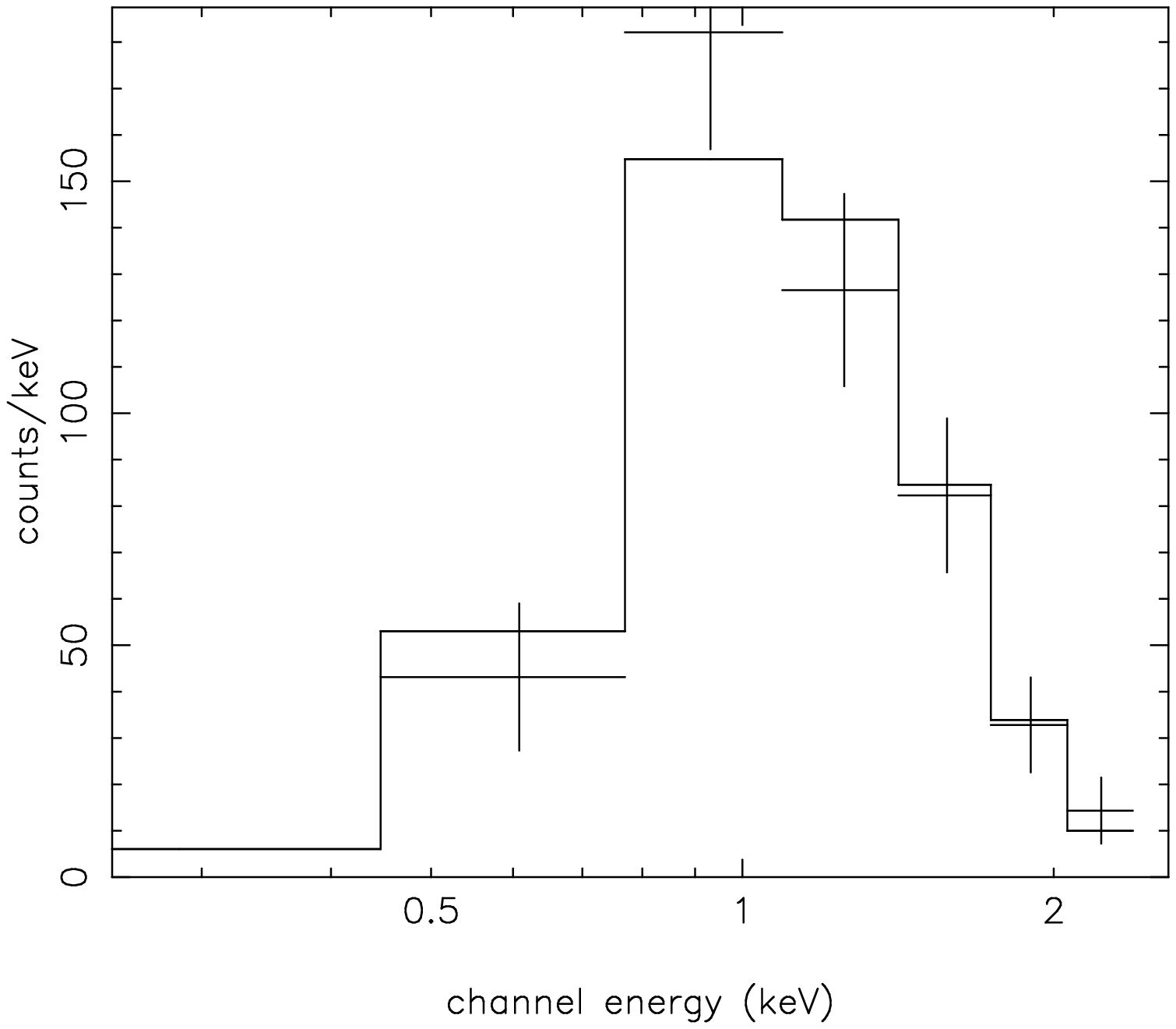}{0.0pt}}
\centerline{\epsfxsize=5truecm 
\figinsert{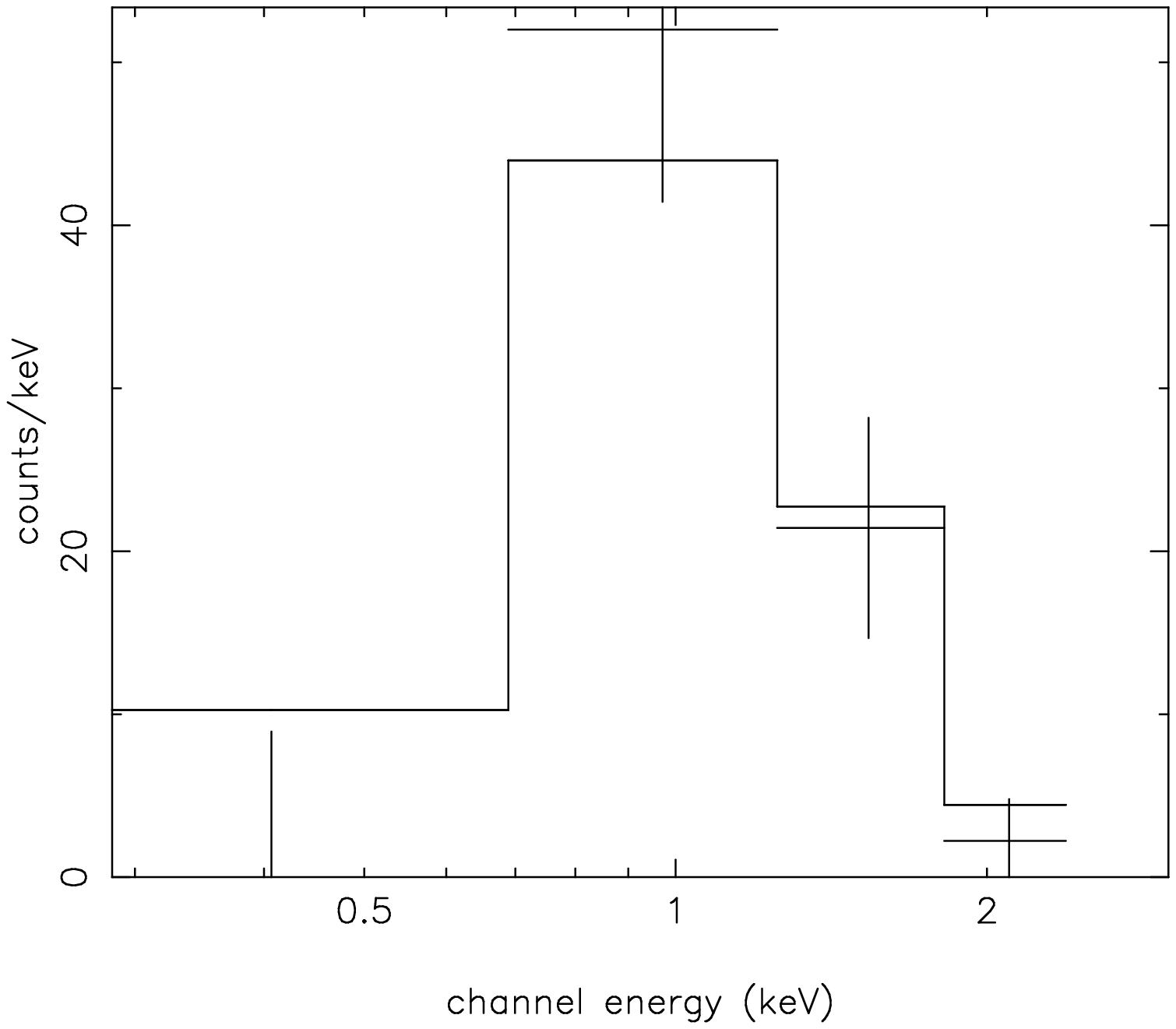}{0.0pt}}
\caption{$0.1-2.4$\,keV  X-ray spectra for the absorbed narrow-emission line galaxies
GSGP4X:091 (a) and GSGP4X:069 (b). The models shown are obtained by fixing the 
intrinsic power law at  $\Gamma=2.2$ and allowing the restframe absorption to vary. }
\end{figure}

\begin {table*}
\begin {center}
\caption {Summary of the 23 probable X-ray emitting galaxies with
$0.1-2.0$\,keV spectral fits for the 9 brightest sources and the stacked spectra of the 
fainter galaxies. The equivalent photon index from a hardness ratio is also quoted
for the faintest galaxies. Optical and X-ray coordinates are given
in B1950 format. A power-law fit is performed with the photoelectric
absorption fixed at the galactic value. The $0.5-2\,$keV flux is given
in erg$\,$s$^{-1}$cm$^{-2}$.  Column 7 gives the offset between the
X-ray source and optical galaxy in arcseconds. Column 10 gives the
$b_j$ magnitude of the galaxy. Note that
$\sim 6$ of these galaxies will be chance associations with X-ray
sources.}
\begin {tabular}{||c|c|c|c|c|c|c|c|c|c|c|c|c||}
Source            & RA$_{x}$    & Dec$_{x}$ & Flux          & $\Gamma$ & $\chi^{2}_{\rm red}$ & $d_{ox}$ & RA$_{o}$ & Dec$_{o}$  & $b_j$   & Em/Abs \\
\hline
$\rm GSGP4X:091 $ &  00 55 36.8 & -28 09 15 & $4.0\times10^{-14}$ &   0.14$\pm$0.30 & 2.77    & 10.4  &  00 55 36.3 & -28 09 23  & 21.33   &  Em \\ 
$\rm GSGP4X:017 $ &  00 54 08.7 & -27 46 23 & $3.7\times10^{-14}$ &   2.25$\pm$0.09 & 1.24    & 17.2  &  00 54 08.5 & -27 46 06  & 18.27   & Abs \\ 
$\rm QSF1X:020  $ &  03 39 41.5 & -45 21 07 & $2.8\times10^{-14}$ &   1.95$\pm$0.21 & 0.33    & 14.5  &  03 39 42.4 & -45 21 18  & 19.66   & Abs \\ 
$\rm GSGP4X:048 $ &  00 54 51.6 & -27 37 49 & $2.7\times10^{-14}$ &   1.82$\pm$0.15 & 1.83    & 11.3  &  00 54 51.8 & -27 38 00  & 20.37   & Em  \\ 
$\rm SGP3X:006  $ &  00 51 38.6 & -28 40 46 & $2.5\times10^{-14}$ &   1.62$\pm$0.18 & 0.64    & 15.0  &  00 51 38.2 & -28 40 32  & 18.78   & Abs \\ 
$\rm QSF1X:036  $ &  03 40 08.3 & -44 48 14 & $2.0\times10^{-14}$ &   2.49$\pm$0.18 & 0.67    & 10.9  &  03 40 07.9 & -44 48 24  & 21.07   & Em \\ 
$\rm GSGP4X:069 $ &  00 55 12.2 & -27 49 19 & $1.3\times10^{-14}$ &   0.79$\pm$0.51 & 3.67    & 9.4   &  00 55 11.5 & -27 49 17  & 20.25   & Em \\ 
$\rm GSGP4X:064 $ &  00 55 07.1 & -27 39 30 & $1.2\times10^{-14}$ &   1.73$\pm$0.35 & 0.65    & 9.5   &  00 55 06.4 & -27 39 32  & 17.84   & Abs  \\ 
$\rm SGP3X:033  $ &  00 52 36.3 & -28 54 37 & $1.0\times10^{-14}$ &   1.94$\pm$0.24 & 0.63    & 17.0  &  00 52 35.2 & -28 54 46  & 18.86   & Em \\ 
$\rm Faint (14) $ &      -      &     -     & $< 1.0\times10^{-14}$ &   1.68$\pm$0.11 & 0.83  & -     &   -         & -          & -       & -  \\ 
$\rm GSGP4X:114 $ &  00 56 11.5 & -28 04 41 & $8.6\times10^{-15}$ &  1.22$\pm$0.51  * & -       & 19.0  &  00 56 11.4 & -28 05 00  & 18.93 &  Em \\ 
$\rm GSGP4X:094 $ &  00 55 42.0 & -27 50 40 & $7.7\times10^{-15}$ &  2.55$\pm$0.48  * & -       & 8.1   &  00 55 41.9 & -27 50 48  & 20.22 &  Em \\ 
$\rm SGP2X:060  $ &  00 50 25.2 & -29 13 38 & $7.4\times10^{-15}$ &  1.44$\pm$0.65  * & -       & 9.0   &  00 50 25.2 & -29 13 29  & 21.27 &  Em \\ 
$\rm SGP2X:025  $ &  00 49 25.9 & -29 20 45 & $6.3\times10^{-15}$ &  $-$0.2$\pm$0.82 * & -       & 9.1  &  00 49 25.8 & -29 20 36  & 21.06 &  Em \\ 
$\rm SGP2X:049  $ &  00 49 51.9 & -29 25 43 & $6.1\times10^{-15}$ &  0.02$\pm$0.90 * & -       & 17.8  &  00 49 52.5 & -29 25 59  & 20.67  &  Abs \\ 
$\rm QSF1X:064  $ &  03 41 26.7 & -45 04 44 & $5.4\times10^{-15}$ &  0.15$\pm$1.10 * & -       & 16.1  &  03 41 27.6 & -45 04 31  & 20.20  &  Em \\ 
$\rm GSGP4X:109 $ &  00 55 59.7 & -27 45 27 & $5.3\times10^{-15}$ &  0.77$\pm$1.15   * & -       &  8.8  &  00 55 59.3 & -27 45 34 & 16.78 &  Em\\ 
$\rm GSGP4X:086 $ &  00 55 31.1 & -27 40 45 & $5.3\times10^{-15}$ &   1.72$\pm$0.82* & -       &  7.1  &  00 55 31.2 & -27 40 38  & 17.94  &  Abs \\ 
$\rm QSF3X:039  $ &  03 40 30.1 & -44 27 15 & $5.2\times10^{-15}$ &   2.57$\pm$0.95* & -       &  6.4  &  03 40 29.5 & -44 27 15  & 19.57  &  Abs\\ 
$\rm GSGP4X:020 $ &  00 54 16.0 & -28 03 12 & $4.8\times10^{-15}$ &   2.95$\pm$1.12* & -       &  7.9  &  00 54 16.6 & -28 03 12  & 21.43  &  Abs\\ 
$\rm QSF1X:033  $ &  03 40 08.5 & -45 10 17 & $4.7\times10^{-15}$ &   0.65$\pm$1.62* & -       & 19.3  &  03 40 10.3 & -45 10 20  & 19.57  &  Em  \\ 
$\rm GSGP4X:082 $ &  00 55 24.6 & -27 39 27 & $4.1\times10^{-15}$ &   0.59$\pm$1.54* & -       &  9.1  &  00 55 24.5 & -27 39 18  & 18.88  &  Em \\ 
$\rm GSGP4X:054 $ &  00 54 55.5 & -28 01 06 & $2.9\times10^{-15}$ &   2.69$\pm$1.61 * & -       &  8.0  &  00 54 55.9 & -28 01 00  & 19.96 &  Em \\ 
$\rm GSGP4X:057 $ &  00 54 59.3 & -27 59 48 & $2.1\times10^{-15}$ &   0.62$\pm$2.30* & -       & 20.0  &  00 55 00.8 & -27 59 50  & 19.52  &  Em \\ 
\hline
\end{tabular}
\end{center}
\noindent{* Equivalent photon index from the $0.5-2.0\,$keV hardness ratio}
\end{table*}

\medskip

\section{Summary and conclusions}

Using a sample of over 300 X-ray sources detected on 5 deep (21-49ks)
$\em ROSAT$ fields we investigate the X-ray spectra of the source
population. Using a hardness ratio analysis we confirm recent claims
that the average source spectra harden towards fainter fluxes from an
equivalent photon index of $\Gamma=2.2$ at
$S_{(0.5-2keV)}=1\times10^{-13}$erg$\,$s$^{-1}$cm$^{-2}$ to
$\Gamma\simeq1.7$ below $1\times10^{-14}$erg$\,$s$^{-1}$cm$^{-2}$. We
then attempt to show the type of source responsible for this trend.
So far 128 QSOs have been identified from this survey and these
dominate the source counts at X-ray fluxes above
$1\times10^{-14}$erg$\,$s$^{-1}$cm$^{-2}$. At fainter fluxes the X-ray
population remains largely unidentified.  We find that the
unidentified sources have harder mean X-ray spectra than QSOs,
regardless of source intensity. We also show that the QSOs detected so
far show no evidence for spectral hardening with decreasing flux,
implying that the change in mean spectra is due to the emergence of
another source population. Recent work has suggested that many of
these are X-ray luminous galaxies (Roche et al 1995a, Boyle et al
1995a, Carballo et al 1995, McHardy et al 1995).  
Taking a subset of 23 X-ray sources
identified as the most likely galaxy candidates, we find that these
show a mean spectral index of $\Gamma=1.69$ from $0.1-2.0$\,keV,
confirming the findings of Carballo et al 1995.
These galaxies consist of 8 absorption-line galaxies and 15 with
emission line features. Hardness ratios suggest that the emission-line
galaxies have significantly harder X-ray spectra than QSOs.  Stacking
the spectra of these faint sources, the absorption-line galaxies yield
a spectral index of $\Gamma=1.94\pm0.08$ while the emission-line
galaxies give $\Gamma=1.51\pm0.09$ from $0.1-2.0$\,keV, more
consistent with the residual X-ray background.  Individually, the
galaxies show a range of spectral properties from very hard X-ray
sources to those with soft, QSO-like spectra. At least 2 emission-line
galaxies show evidence for significant photoelectric absorption in the
range $N_H\sim10^{21}-10^{22}\,$atom$\,$cm$^{-2}$.  Given recent
results which suggest that the surface density of narrow emission-line
galaxies could approach that of AGN at
$\sim1\times10^{-15}$erg$\,$s$^{-1}$cm$^{-2}$ and possibly contribute
$15-40\%$ of the total XRB at 1keV (Boyle et al 1995a, Griffiths et al
1995b), these spectral results provide further evidence that they may
be the missing component of the cosmic XRB.

\section*{ACKNOWLEDGMENTS}

OA was funded by an SERC/PPARC studentship. BJB was partly supported
by a Royal Society University Research Fellowship during part of this
work.  OA thanks Andy Fabian for helpful advice and discussions. We
also thank the staff at the Anglo-Australian observatory and the $\em
ROSAT$ team for making these observations possible.

\end{document}

%% file: epsf_mn.tex
\newread\epsffilein    
\newif\ifepsffileok    
\newif\ifepsfbbfound   
\newif\ifepsfverbose   
\newdimen\epsfxsize    
\newdimen\epsfysize    
\newdimen\epsftsize    
\newdimen\epsfrsize    
\newdimen\epsftmp      
\newdimen\pspoints     
\def\epsfbox#1{\global\def\epsfllx{72}\global\def\epsflly{72}%
   \global\def\epsfurx{540}\global\def\epsfury{720}%
   \def\lbracket{[}\def\testit{#1}\ifx\testit\lbracket
   \let\next=\epsfgetlitbb\else\let\next=\epsfnormal\fi\next{#1}}%
\def\epsfgetlitbb#1#2 #3 #4 #5]#6{\epsfgrab #2 #3 #4 #5 .\\%
   \epsfsetgraph{#6}}%
\def\epsfnormal#1{\epsfgetbb{#1}\epsfsetgraph{#1}}%
\def\epsfgetbb#1{%
\openin\epsffilein=#1
\ifeof\epsffilein\errmessage{I couldn't open #1, will ignore it}\else
   {\epsffileoktrue \chardef\other=12
    \def\do##1{\catcode`##1=\other}\dospecials \catcode`\ =10
    \loop
       \read\epsffilein to \epsffileline
       \ifeof\epsffilein\epsffileokfalse\else
          \expandafter\epsfaux\epsffileline:. \\%
       \fi
   \ifepsffileok\repeat
   \ifepsfbbfound\else
    \ifepsfverbose\message{No bounding box comment in #1; using defaults}\fi\fi
   }\closein\epsffilein\fi}%
\def\epsfsetgraph#1{%
   \epsfrsize=\epsfury\pspoints
   \advance\epsfrsize by-\epsflly\pspoints
   \epsftsize=\epsfurx\pspoints
   \advance\epsftsize by-\epsfllx\pspoints
   \epsfxsize\epsfsize\epsftsize\epsfrsize
   \ifnum\epsfxsize=0 \ifnum\epsfysize=0
      \epsfxsize=\epsftsize \epsfysize=\epsfrsize
     \else\epsftmp=\epsftsize \divide\epsftmp\epsfrsize
       \epsfxsize=\epsfysize \multiply\epsfxsize\epsftmp
       \multiply\epsftmp\epsfrsize \advance\epsftsize-\epsftmp
       \epsftmp=\epsfysize
       \loop \advance\epsftsize\epsftsize \divide\epsftmp 2
       \ifnum\epsftmp>0
          \ifnum\epsftsize<\epsfrsize\else
             \advance\epsftsize-\epsfrsize \advance\epsfxsize\epsftmp \fi
       \repeat
     \fi
   \else\epsftmp=\epsfrsize \divide\epsftmp\epsftsize
     \epsfysize=\epsfxsize \multiply\epsfysize\epsftmp   
     \multiply\epsftmp\epsftsize \advance\epsfrsize-\epsftmp
     \epsftmp=\epsfxsize
     \loop \advance\epsfrsize\epsfrsize \divide\epsftmp 2
     \ifnum\epsftmp>0
        \ifnum\epsfrsize<\epsftsize\else
           \advance\epsfrsize-\epsftsize \advance\epsfysize\epsftmp \fi
     \repeat     
   \fi
   \ifepsfverbose\message{#1: width=\the\epsfxsize, height=\the\epsfysize}\fi
   \epsftmp=10\epsfxsize \divide\epsftmp\pspoints
   \newcount\figskipcount
      \message{#1 \the\epsfysize  }
   \vbox to\epsfysize{\vfil\hbox to\epsfxsize{%
      \includegraphics{#1}%
      \hfil}}%
\epsfxsize=0pt\epsfysize=0pt}%

{\catcode`\%=12 \global\let\epsfpercent=
\long\def\epsfaux#1#2:#3\\{\ifx#1\epsfpercent
   \def\testit{#2}\ifx\testit\epsfbblit
      \epsfgrab #3 . . . \\%
      \epsffileokfalse
      \global\epsfbbfoundtrue
   \fi\else\ifx#1\par\else\epsffileokfalse\fi\fi}%
\def\epsfgrab #1 #2 #3 #4 #5\\{%
   \global\def\epsfllx{#1}\ifx\epsfllx\empty
      \epsfgrab #2 #3 #4 #5 .\\\else
   \global\def\epsflly{#2}%
   \global\def\epsfurx{#3}\global\def\epsfury{#4}\fi}%
\def\epsfsize#1#2{\epsfxsize}